\documentclass[aps,prd,nofootinbib,twocolumn,superscriptaddress,preprintnumbers]{revtex4}
\usepackage{amssymb}
\usepackage{amsmath}
\usepackage{epsfig}
\usepackage{hyperref}
\usepackage{breakurl}

\makeatletter
\def\simgt{\mathrel{\lower2.5pt\vbox{\lineskip=0pt\baselineskip=0pt
           \hbox{$>$}\hbox{$\sim$}}}}
\def\simlt{\mathrel{\lower2.5pt\vbox{\lineskip=0pt\baselineskip=0pt
           \hbox{$<$}\hbox{$\sim$}}}}
\makeatother

\newcommand{\be}{\begin{equation}}
\newcommand{\ee}{\end{equation}}
\newcommand{\bea}{\begin{eqnarray}}
\newcommand{\eea}{\end{eqnarray}}
\newcommand{\Eq}[1]{Eq.~(\ref{#1})}
\newcommand{\Eqs}[2]{Eqs.~(\ref{#1}) and (\ref{#2})}
\newcommand{\Sec}[1]{Sec.~\ref{#1}}

\newcommand{\feff}{f_{\rm eff}}
\newcommand{\Feff}{F_{\rm eff}}
\newcommand{\MPl}{M_{\rm Pl}}
\newcommand{\sslash}[1]{\ensuremath\raisebox{-0.00cm}{{\small\slash}}\hspace{-0.21cm}#1\/}

\begin{document}

\preprint{UCB-PTH-09/04}
\preprint{MIT-CTP-4116}

\title{Goldstini}

\author{Clifford Cheung}
\affiliation{Berkeley Center for Theoretical Physics, 
  University of California, Berkeley, CA 94720, USA}
\affiliation{Theoretical Physics Group, 
  Lawrence Berkeley National Laboratory, Berkeley, CA 94720, USA}

\author{Yasunori Nomura}
\affiliation{Berkeley Center for Theoretical Physics, 
  University of California, Berkeley, CA 94720, USA}
\affiliation{Theoretical Physics Group, 
  Lawrence Berkeley National Laboratory, Berkeley, CA 94720, USA}
\affiliation{Institute for the Physics and Mathematics of the Universe, 
  University of Tokyo, Kashiwa 277-8568, Japan}

\author{Jesse Thaler}
\affiliation{Center for Theoretical Physics, 
  Massachusetts Institute of Technology, Cambridge, MA 02139, USA}

\begin{abstract}
Supersymmetric phenomenology has been largely bound to the hypothesis 
that supersymmetry breaking originates from a single source.  In this 
paper, we relax this underlying assumption and consider a multiplicity 
of sectors which independently break supersymmetry, thus yielding 
a corresponding multiplicity of goldstini.  While one linear combination 
of goldstini is eaten via the super-Higgs mechanism, the orthogonal 
combinations remain in the spectrum as physical degrees of freedom. 
Interestingly, supergravity effects induce a universal tree-level mass 
for the goldstini which is exactly twice the gravitino mass.  Since 
visible sector fields can couple dominantly to the goldstini rather 
than the gravitino, this framework allows for substantial departures 
from conventional supersymmetric phenomenology.  In fact, this even 
occurs when a conventional mediation scheme is augmented by additional 
supersymmetry breaking sectors which are fully sequestered.  We discuss 
a number of striking collider signatures, including various novel 
decay modes for the lightest observable-sector supersymmetric particle, 
gravitinoless gauge-mediated spectra, and events with multiple displaced 
vertices.  We also describe goldstini cosmology and the possibility 
of goldstini dark matter.
\end{abstract}

\maketitle

\section{Introduction}

Supersymmetry (SUSY) is a theoretically motivated and well-studied 
framework which solves the hierarchy problem and offers a rich 
phenomenology~\cite{Martin:1997ns}.  Of course, if SUSY is to be 
realized in nature, then it must be spontaneously broken.  To this 
end, it is conventionally assumed that SUSY breaking originates 
from the dynamics of a single hidden sector.

While the notion of single sector SUSY breaking is convenient as 
a simplifying premise, it is not very generic in light of top-down 
considerations.  In particular, string theoretic constructions 
routinely predict a multiplicity of geographically sequestered 
sectors~\cite{Giddings:2001yu}, any number of which could independently 
break SUSY.  In this paper we will explore the generic implications 
of multiple sector SUSY breaking.

Consider the low energy effective field theory describing $N$ such 
sequestered sectors.  In the limit in which these sectors are completely 
decoupled---even gravitationally---they enjoy an $N$-fold enhanced 
Poincar\'e symmetry because energy and momentum are {\it separately} 
conserved within each sector.  Likewise, if SUSY is a symmetry of nature 
then it is similarly enhanced, such that
\be
  \textrm{SUSY} \overset{\textrm{decoupled}}{\longrightarrow} 
  \textrm{SUSY}^N \equiv \otimes \prod_{i=1}^N \textrm{SUSY}_i.
\ee
Because this enhancement is an accidental consequence of the decoupling 
limit, gravitational interactions explicitly break SUSY$^N$ down to 
a diagonal combination corresponding to the genuine supergravity (SUGRA) 
symmetry.  Consequently, the ``orthogonal'' SUSY$^{N-1}$ are only 
approximate global symmetries.

In the event that $F$-term breaking occurs independently in each sector, 
each SUSY$_i$ will be spontaneously broken at a scale $F_i$, yielding 
a corresponding goldstino $\eta_i$.%
\footnote{Throughout the paper we take a field basis where $F_i$ are 
 all real and positive, and assume that $F_i \geq F_{i+1}$ without 
 loss of generality.  We will also focus on the case where SUSY breaking 
 still occurs in the $\MPl \rightarrow \infty$ limit, and only briefly 
 comment on ``almost no-scale'' SUSY-breaking sectors in the Appendix. 
 The possibility of $D$-term breaking will be left to future work.}
In unitary gauge, one linear combination of goldstini, $\eta_{\rm long}$, 
is eaten by the gravitino via the super-Higgs mechanism, leaving $N-1$ 
goldstini in the spectrum.  We denote these fields by $\zeta_a$, where 
$a=1,\ldots,N-1$.

Since the remaining $N-1$ goldstini correspond to the approximate 
SUSY$^{N-1}$ which are explicitly broken by SUGRA, one should not 
expect these goldstini to remain exactly massless.  In fact, we will 
show that they acquire a {\it tree-level} mass
\be
  m_a = 2 m_{3/2},
\label{eq:keyequation}
\ee
induced by gravitational effects.  As we will see, the curious factor 
of $2$ is ultimately fixed by the symmetries of SUGRA, and we will 
robustly derive it in a number of different ways.

Up to now, SUSY phenomenology has been almost exclusively devoted to 
a scenario in which the gravitino and the goldstino are effectively 
one and the same.%
\footnote{To our knowledge, the only mention of multiple goldstini 
 in the literature appears in Ref.~\cite{Benakli:2007zza}.}
In the context of multiple sector SUSY breaking, however, this corresponds 
to a rather privileged arrangement in which the dominant contributions 
to SUSY breaking in the supersymmetric standard model (SSM) sector arise 
from the SUSY breaking sector with the highest SUSY breaking scale.  In 
any other situation, the SSM fields will actually couple more strongly 
to the goldstini than to the gravitino, and this will have a significant 
impact on collider physics and cosmology.  A simple context in which 
this occurs is when a conventional SUSY breaking scenario is augmented 
by additional SUSY breaking sectors which are fully sequestered 
(see Fig.~\ref{fig:setup-1} in \Sec{subsec:single-med}).

This paper is organized as follows.  In \Sec{sec:goldstone}, we review 
an analogous construction for Goldstone bosons arising from multiple 
symmetry breaking.  The goldstini case of multiple SUSY breaking 
is then presented in \Sec{sec:goldstiniFC}.  We derive the relation 
$m_a = 2 m_{3/2}$ in \Sec{sec:masses}, using both a St\"{u}ckelberg 
method and a conformal compensator method.  A direct SUGRA calculation 
of the factor of two appears in the Appendix.  Corrections to this 
mass relation are given in \Sec{sec:correction}, and the couplings 
to the SSM are given in \Sec{sec:vis-int}.  Possible LHC signatures 
of this scenario---including wrong mass ``gravitinos'', gravitinoless 
gauge mediation, smoking gun evidence for the factor of two, 
three-body neutralino decays, and displaced monojets---are presented 
in \Sec{sec:collider}.  Goldstini cosmology is described in 
\Sec{sec:cosmology}, including scenarios that yield goldstini 
dark matter.  We conclude in \Sec{sec:discuss}.

\section{Goldstone Analogy}
\label{sec:goldstone}

Because the notion of multiple sector SUSY breaking is not a familiar 
one, it is instructive to analyze an analogous construction involving 
multiple $U(1)$ symmetry breaking.  Consider a scenario in which $\phi_1$ 
and $\phi_2$ are complex scalar fields which enjoy separate global 
symmetries $U(1)_1$ and $U(1)_2$.  Furthermore, assume that the diagonal 
$U(1)_V$ is gauged and that $\phi_1$ and $\phi_2$ have no direct couplings 
except for gauge interactions.

\subsection{Fields and Couplings}
\label{subsec:analogyfields}

If $\phi_1$ and $\phi_2$ separately acquire vacuum expectation values 
(vevs), then we can non-linearly parameterize the Goldstone modes as
\be
  \phi_i = f_i e^{i \pi_i/\sqrt{2} f_i},
\label{eq:Goldstone-1}
\ee
for $i=1,2$.  One linear combination of $\pi_1$ and $\pi_2$ is eaten 
via the Higgs mechanism.  The orthogonal combination, $\varphi$, 
corresponds to a physical pseudo-Goldstone boson that arises from 
the spontaneous breaking of a global $U(1)_A$ axial symmetry. 
Concretely, go to a basis
\bea
  \left( \begin{array}{c}
    \pi_1 \\ \pi_2
  \end{array} \right)
&=&
  \left( \begin{array}{cc}
    \cos\theta & -\sin\theta \\
    \sin\theta &  \cos\theta
  \end{array} \right)
  \left( \begin{array}{c}
    \pi_{\rm long} \\ \varphi
  \end{array} \right)
\nonumber\\
  &\overset{\textrm{unitary}}{\overset{\textrm{gauge}}{\longrightarrow}}& 
  \left( \begin{array}{c}
    -\sin\theta \, \varphi \\ \cos\theta \, \varphi
\end{array} \right),
\label{eq:Goldstone-2}
\eea
where $\tan \theta = f_2/f_1$ and $\feff = \sqrt{f_1^2+f_2^2}$.  In unitary 
gauge, $\pi_{\rm long}$ becomes the longitudinal mode of the $U(1)_V$ 
gauge boson.

The interactions of $\varphi$ with other fields can be obtained from 
plugging the parameterization of \Eqs{eq:Goldstone-1}{eq:Goldstone-2} 
into couplings involving $\phi_i$ and those fields.  Note a crucial 
difference between the couplings of $\pi_{\rm long}$ and $\varphi$. 
While one can always do field redefinitions such that $\pi_{\rm long}$ 
couples only derivatively, ${\cal L}_{\rm int} = (1/\feff) 
(\partial_\mu \pi_{\rm long}) J^\mu$ where $J^\mu$ is the $U(1)_V$ 
current, there is no guarantee that the same can be done for $\varphi$.

\subsection{Masses}
\label{subsec:analogymasses}

As is well known, $\pi_1$ and $\pi_2$ are exactly massless in the limit 
in which $U(1)_1 \times U(1)_2$ is an exact symmetry of the Lagrangian. 
One way of understanding this fact is to consider the unitary gauge 
Lagrangian for the massive $U(1)_V$ gauge boson,
\be
  {\cal L}_{\rm unit} 
  = -\frac{1}{4g^2}F_{\mu\nu}F^{\mu\nu} - f^2 A_\mu A^\mu.
\ee
For the moment, let us assume that $U(1)_1$ is broken but $U(1)_2$ is 
preserved.  As a consequence, there is a single eaten Goldstone mode, 
$\pi_1$.  Using the St\"{u}ckelberg replacement, we can reinstate $\pi_1$ 
as a propagating degree of freedom by applying a gauge transformation
\be
  A_\mu \rightarrow A_\mu + \frac{1}{\sqrt{2} f} \partial_\mu \pi_1,
\ee
and promoting $\pi_1$ to a dynamical field.  Doing so yields
\be
  {\cal L} = -\frac{1}{2} \partial_\mu \pi_1 \partial^\mu \pi_1 
    + \mbox{terms involving $A_\mu$}.
\ee
Obviously, the exact same argument can apply in the case in which $U(1)_2$ 
is broken and $U(1)_1$ is preserved.  Thus, if $U(1)_1$ and $U(1)_2$ are 
independently broken, then the Lagrangian must take the form
\be
  {\cal L} = -\frac{1}{2} \sum_i \partial_\mu \pi_i \partial^\mu \pi_i 
    + \mbox{terms involving $A_\mu$},
\ee
and so a mass term is forbidden for either $\pi_i$.  Said another way, 
either $\pi_i$ could have been eaten by $A_\mu$, so both are required 
to be massless.  This implies that the uneaten Goldstone mode, $\varphi$, 
is massless.

If it is not the case that $U(1)_1 \times U(1)_2$ is an exact symmetry, 
then the above argument is only approximate.  In particular, any explicit 
$U(1)_A$ violating, $U(1)_V$ preserving operators will provide a mass 
term for the uneaten mode, $\varphi$, at tree level.  Moreover, even 
if such operators are missing, they can be generated radiatively.  For 
example, this occurs in a non-Abelian Goldstone theory in which 
$\phi_1$ and $\phi_2$ are in fundamental representations of $SU(k)_1$ 
and $SU(k)_2$ global symmetries, respectively, of which the diagonal 
$SU(k)_V$ combination is gauged.  Since the gauge interactions explicitly 
violate the $SU(k)_A$ global symmetry, radiative corrections will 
generate operators of the form
\be
  |\phi_1^\dagger \phi_2|^2,
\label{eq:explicitbreak}
\ee
which induce a mass for $\varphi$, albeit at loop level.  As we will see 
shortly, the non-Abelian theory provides the closest analogy to multiple 
sector SUSY breaking---SUGRA, which is precisely the gauged diagonal 
SUSY, explicitly violates the orthogonal SUSY$^{N-1}$ and thus induces 
nonzero masses for the uneaten goldstini.  The important difference in
the case of SUSY is that these masses will arise at tree level rather 
than at loop level.

\section{Goldstini Fields and Couplings}
\label{sec:goldstiniFC}

The discussion of multiple sector SUSY breaking exactly parallels 
that of the previous section.  We will focus here on the case of 
$F$-term breaking, and imagine that there exist two chiral superfields, 
$X_1$ and $X_2$, that reside in two sequestered sectors.  In the absence 
of direct couplings, gravitational or otherwise, these fields enjoy an 
enhanced $\textrm{SUSY}_1 \otimes \textrm{SUSY}_2$ symmetry.  Assuming 
that the highest component of $X_i$ acquires a vev equal to $F_i$, then 
SUSY$_i$ is broken and we can use the non-linear parameterization
\bea
  X_i &=& e^{Q \eta_i/\sqrt{2}F_i} (x_i +\theta^2 F_i)
\nonumber\\
  &=&  x_i +\eta_i^2/2F_i +\sqrt{2} \theta \eta_i + \theta^2 F_i,
\label{eq:nonlinparam}
\eea
for $i=1,2$, where $Q = \partial / \partial \theta$ is the generator of 
SUSY transformations and we have neglected all derivatively coupled terms.%
\footnote{A similar non-linear parametrization was considered in 
 Ref.~\cite{Komargodski:2009rz} for a single goldstino.}
Here $\eta_i$ is the goldstino corresponding to the $F$-term breaking 
of SUSY$_i$.  Note that this form is identical to the usual linear 
parameterization of a chiral superfield except for the presence of 
$\eta_i^2$ in the lowest component of $X_i$.

In the presence of SUGRA, the diagonal combination of SUSY$_1$ and 
SUSY$_2$ is gauged, and thus one of the goldstini is eaten.  As before, 
it is convenient to work in a basis
\bea
  \left( \begin{array}{c}
    \eta_1 \\ \eta_2
  \end{array} \right)
&=&
  \left( \begin{array}{cc}
    \cos\theta & -\sin\theta \\
    \sin\theta &  \cos\theta
  \end{array} \right)
  \left( \begin{array}{c}
    \eta_{\rm long} \\ \zeta
  \end{array} \right)
\nonumber\\
  &\overset{\textrm{unitary}}{\overset{\textrm{gauge}}{\longrightarrow}}& 
  \left( \begin{array}{c}
    -\sin\theta \, \zeta \\ \cos\theta \, \zeta
  \end{array} \right),
\label{eq:goldstinibasis}
\eea
where $\tan\theta = F_2/F_1$ and $\Feff = \sqrt{F_1^2+F_2^2}$. 
Thus, $\eta_{\rm long}$ is eaten by the gravitino, and $\zeta$ remains 
a propagating degree of freedom.

The interactions of $\zeta$ with other fields can be obtained using 
the parameterization of \Eqs{eq:nonlinparam}{eq:goldstinibasis}.  Since 
$X_i$ is a true chiral superfield, the couplings of $\zeta$ can be obtained 
directly in superspace.  While one can always work in a field basis 
where $\eta_{\rm long}$ couples only derivatively, ${\cal L}_{\rm int} 
= (1/\Feff) (\partial_\mu \eta_{\rm long}) \tilde{J}^\mu$ where 
$\tilde{J}^\mu$ is the supercurrent, the same cannot be done in 
general for $\zeta$.

If the number of sequestered SUSY breaking sectors is greater than two, 
then there will be multiple uneaten goldstini $\zeta_a$, which are related 
to $\eta_i$ by
\be
  \eta_i = V_{ia} \zeta_a,
\label{eq:V_ia}
\ee
where $V_{ia}$ is the $N \times (N-1)$ part of the unitary matrix which 
goes from the $\eta_i$ basis to the $\{ \eta_{\rm long}, \zeta_a \}$ 
basis.  The $\zeta_a$ fields are orthogonal to the eaten mode.  Since
\be
  \eta_{\rm long} = \frac{1}{\Feff} \sum_i F_i \eta_i,
\ee
this implies $\sum_i F_i V_{ia} = 0$.  The form of $V_{ia}$ is determined 
by the mass matrix of $\zeta_a$, which we will now discuss.

\section{Goldstini Masses}
\label{sec:masses}

In \Sec{subsec:analogymasses}, we saw that uneaten Goldstone bosons 
typically acquire masses from loops of non-Abelian gauge bosons.  SUGRA 
effects similarly induce masses for the goldstini---only this happens 
at tree level!  More precisely, in the limit in which each sector couples 
only through SUGRA, all goldstini acquire a tree level mass which is 
universal and given by $m_a = 2 m_{3/2}$.%
\footnote{It may appear contradictory that the goldstini acquire 
 a tree-level mass, since they are derivatively coupled in the limit 
 of global SUSY.  Nevertheless, for finite $\MPl$, the goldstini couple 
 not just as $\partial_\mu \eta_i$ but also as $\sigma_\mu \bar{\eta}_i$, 
 so a mass term is not forbidden.}
While the factor of $2$ may be verified explicitly by considering the 
explicit SUGRA Lagrangian (see the Appendix), we find it more illuminating 
to derive it in two separate but more direct ways.  Collider and 
cosmological implications of this universal mass will be discussed 
in Secs. \ref{sec:collider} and \ref{sec:cosmology}.

\subsection{Two via St\"{u}ckelberg}
\label{subsec:2-stuckelberg}

The simplest way of understanding $m_a = 2 m_{3/2}$ is in analogy 
with the logic of \Sec{subsec:analogymasses}.  We start from a unitary 
gauge SUGRA Lagrangian, where the quadratic action for the gravitino 
is~\cite{Wess:1992cp}
\be
  {\cal L}_{\rm unit} = \epsilon^{\mu\nu\rho\sigma} 
    \bar{\psi}_\mu \bar{\sigma}_\nu \partial_\rho \psi_\sigma 
    - m_{3/2} \left( \psi_\mu \sigma^{\mu\nu} \psi_\nu 
    + \bar{\psi}_\mu \bar{\sigma}^{\mu\nu} \bar{\psi}_\nu \right),
\ee
where $\sigma^{\mu\nu} \equiv (\sigma^\mu \bar{\sigma}^\nu - \sigma^\nu 
\bar{\sigma}^\mu)/4$ and $m_{3/2} \simeq \Feff / \sqrt{3} \MPl$.%
\footnote{In this paper we assume that SUSY is broken in the global limit 
 and that the SUSY breaking vacuum is unaffected by finite $\MPl$ effects. 
 Some of the equations below, e.g.\ \Eq{eq:SUGRAtrans}, do not hold if 
 we relax this assumption (see Ref.~\cite{ArkaniHamed:2004yi} for a clear 
 discussion of the general SUSY transformation laws for the gravitino). 
 A more general case is discussed briefly in the Appendix.}
Now consider the scenario in which SUSY has been broken {\it only} in 
sector~$1$, and the corresponding goldstino $\eta_1$ has been eaten. 
We can reinstate the $\eta_1$ degree of freedom by applying the 
St\"{u}ckelberg construction---that is, applying a SUGRA transformation 
on the unitary gauge Lagrangian, and then promoting the SUGRA 
parameter to a dynamical field.  In particular, we apply the 
SUGRA transformation~\cite{Wess:1992cp}
\be
  \psi_\mu \rightarrow \psi_\mu 
    + \sqrt{\frac{2}{3}} m_{3/2}^{-1} \partial_\mu \eta_1 
    + \frac{i}{\sqrt{6}} \sigma_\mu \bar{\eta}_1,
\label{eq:SUGRAtrans}
\ee
to the Lagrangian, yielding
\be
  {\cal L} = -i \bar{\eta}_1 \bar{\sigma}^\mu \partial_\mu \eta_1 
    - \frac{1}{2} (2 m_{3/2}) (\eta_1^2 +\bar{\eta}_1^2) + \ldots,
\label{eq:stuckel1}
\ee
where the ellipses denote all terms involving the $\psi_\mu$, including 
the mixing terms between the gravitino and the goldstino.  Note how the 
kinetic term for $\eta_1$ is generated by the cross term obtained from 
\Eq{eq:SUGRAtrans}.  Given the normalization of a Majorana fermion, this 
implies a goldstino mass of $m_1 = 2 m_{3/2}$.

Now of course if there is only one goldstino, then this mass is not 
physical, since $\eta_1$ is eaten via the super-Higgs mechanism.  However, 
if there is multiple sector SUSY breaking, then there will be several 
goldstini $\eta_i$.  Since any one of the $X_i$ {\it could} have broken 
SUSY on its own and been eaten by the gravitino, all of them must 
take the form of \Eq{eq:stuckel1}.  Thus, with multiple goldstini, 
the St\"{u}ckelberg Lagrangian becomes
\be
  {\cal L} = \sum_i 
    \left\{ -i \bar{\eta}_i \bar{\sigma}^\mu \partial_\mu \eta_i 
    - \frac{1}{2} (2 m_{3/2}) (\eta_i^2 +\bar \eta_i^2) \right\} + \ldots,
\label{eq:stuckel}
\ee
where the ellipses include the mixing between the gravitino and the 
eaten goldstino, which is now some linear combination of the $\eta_i$.%
\footnote{One might think that each $\eta_i$ should have a mass given 
 by $2 F_i/\sqrt{3} \MPl$ (twice the gravitino mass for sector $i$ alone) 
 instead of $2 \Feff/\sqrt{3} \MPl$ (twice the gravitino mass for all 
 sectors together).  This, however, would not lead to the correct mass 
 for the eaten mode, which must take the form of \Eq{eq:stuckel1}.}
We can now rotate the fermions by an orthogonal matrix, to isolate the 
eaten goldstino mode, and then go to unitary gauge.  Since the $\eta_i$ 
mass matrix is proportional to the identity, the leftover goldstini, 
$\zeta_a$, will all have mass $m_a = 2 m_{3/2}$.

\subsection{Two via the Conformal Compensator}
\label{subsec:2-compensator}

An alternative way of understanding the relation $m_a = 2 m_{3/2}$ 
is to use the conformal compensator formalism~\cite{Cremmer:1978hn}. 
Morally, the factor of $2$ arises because the conformal compensator 
couples to mass dimension, and $F_i$ has mass dimension $2$.  To see 
this in a simple example, consider the case of several sectors which 
independently break SUSY via a Polonyi superpotential
\bea
  {\cal L} &=& \int\! d^4\theta\, 
    C^\dagger C \sum_i ( X_i^\dagger X_i + \ldots )
\nonumber\\
  && {} + \int\! d^2\theta\, C^3 \sum_i \mu_i^2 X_i + \textrm{h.c.},
\eea
where $C = 1 + \theta^2 m_{3/2}$ is the conformal compensator, the dots 
indicate higher order terms necessary to stabilize the scalar components 
of $X_i$, and we have chosen a sequestered form for the K\"{a}hler 
potential.  By rescaling $X_i \rightarrow X_i/C$, we see that $C$ only 
couples to dimensionful parameters---namely, $\mu_i$.  Plugging in for 
the lowest component of the non-linear parameterization of $X_i$ in 
\Eq{eq:nonlinparam}, we obtain
\bea
  {\cal L} &\supset& \int\! d^2\theta\, C^2 \sum_i \mu_i^2 X_i
\nonumber\\
  &=& {} - \frac{1}{2} (2 m_{3/2}) \sum_i \eta_i^2 + \textrm{const.},
\eea
where we have solved for the auxiliary fields $F_i = -\mu_i^2$ and plugged 
in for the conformal compensator.  The fact that $\mu_i^2$ has mass 
dimension $2$ implies that conformal compensator couples as $C^2$, 
yielding the important factor of $2$ in the goldstini mass.

\section{Deviations from the Sequestered Limit}
\label{sec:correction}

So far, we have limited our discussion to the case where the only 
interactions between SUSY breaking sectors arise from SUGRA.  This 
is certainly the case if every sector, including the SSM sector, is 
sequestered from one another and SUSY breaking is communicated to 
the SSM via SUGRA effects, i.e.\ anomaly mediation.  In this section, 
we consider the case where one or more SUSY breaking sectors have 
direct couplings to the SSM to mediate SUSY breaking.  We discuss 
effects of such couplings on the goldstini properties.

\subsection{Single Sector Mediation}
\label{subsec:single-med}

\begin{figure}
\begin{center}
\includegraphics[scale=0.75]{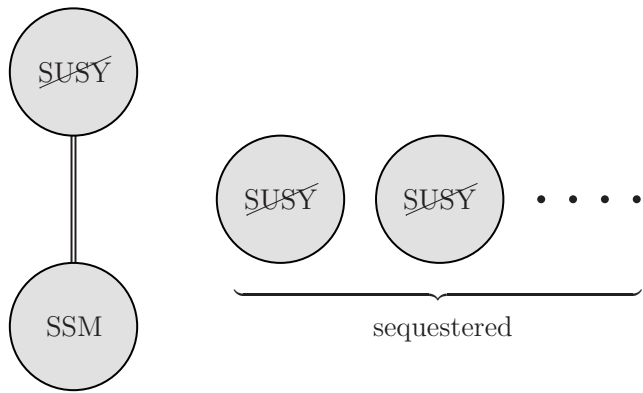}
\end{center}
\caption{A schematic depiction of a scenario in which the SSM sector 
 couples to only one of the SUSY breaking sectors.  Note that this setup 
 still leads to interactions between SSM fields and goldstini in the 
 sequestered sectors, since the goldstino of the sector coupling to the 
 SSM is in general a linear combination of the gravitino and uneaten 
 goldstini.}
\label{fig:setup-1}
\end{figure}
The simplest deviation from the fully sequestered limit is for direct 
couplings to exist only between the SSM and {\it one} of the SUSY breaking 
sectors, as illustrated in Fig.~\ref{fig:setup-1}.  This corresponds to 
the situation where a conventional SUSY breaking scenario, such as gauge 
mediation, is augmented by one or more fully-sequestered SUSY breaking 
sectors.  This may easily arise in realistic top-down setups.

Despite the coupling to the SSM, the different SUSY breaking sectors 
themselves still interact only through SUGRA, so the analysis of the 
goldstini masses in the previous sections remains intact.  Note that 
because of the mixing matrix from \Eq{eq:V_ia}, there are still nontrivial 
couplings between SSM fields and the goldstini from the sequestered SUSY 
breaking sectors.

\subsection{Induced Couplings between SUSY Breaking Sectors}
\label{subsec:induced-coupl}

\begin{figure}
\begin{center}
\includegraphics[scale=0.75]{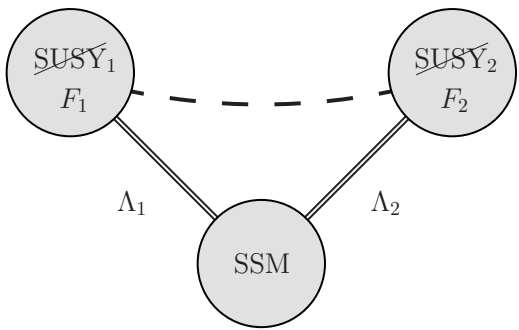}
\end{center}
\caption{A schematic depiction of a scenario in which sectors 1 and 2 
 have direct interactions to the SSM sector via operators suppressed 
 by $\Lambda_1$ and $\Lambda_2$, respectively (double lines).  These 
 interactions induce direct couplings between sectors 1 and 2 through 
 radiative corrections (dashed line).}
\label{fig:setup-2}
\end{figure}
If two or more SUSY breaking sectors have direct couplings to the SSM, 
a true deviation from the sequestered limit arises.  To see how this 
happens, consider sectors 1 and 2, each of which couples to the SSM 
sector via an operator suppressed by $\Lambda_1$ and $\Lambda_2$, 
respectively (see Fig.~\ref{fig:setup-2}).  Clearly, loops of SSM 
sector fields induce direct interactions between sectors 1 and 2, 
which may in turn modify the goldstini properties.

Direct interactions between SUSY breaking sectors can potentially modify 
the vacuum structure drastically so that SUSY breaking no longer occurs 
in some of these sectors.  We assume that this is not the case, i.e.\ 
parameters take values such that the shift of the vacuum is small 
enough to preserve the essential structure of the sectors.  (The parameter 
regions considered in later sections satisfy this condition.)  It 
is then easiest to analyze the effect of direct couplings using the 
non-linear parameterization of \Eq{eq:nonlinparam}, where $x_i$ and 
$F_i$ represent the values after the vacuum shift and $\eta_i$ is the 
goldstino arising from sector~$i$.

Since $\eta_1$ and $\eta_2$ have the quantum numbers conjugate to the 
generators of SUSY$_1$ and SUSY$_2$, respectively, they have a unit 
charge under the corresponding $R$ symmetries, $U(1)_{R_1}$ and 
$U(1)_{R_2}$, rotating these generators.  Consequently, any deviations 
of the goldstini Majorana masses from $2m_{3/2}$ require an additional 
$R$-symmetry breaking transmission between sectors 1 and 2 beyond 
that provided by SUGRA through $m_{3/2}$.  Since the setup considered 
here has tree-level direct couplings only between the SSM and SUSY 
breaking sectors, such a transmission must occur through the SSM sector.

The leading $R$-breaking transmitting couplings between a SUSY 
breaking sector and the SSM sector are given by the gaugino-mass 
and $A$-term operators, $\int\! d^2\theta\, X_i {\cal W}^\alpha 
{\cal W}_\alpha/\Lambda_i$ and $\int\! d^2\theta\, X_i \Phi^\dagger 
\Phi/\Lambda_i$, which may or may not exist depending on the properties 
of the SUSY breaking sector.  Here, ${\cal W}_\alpha$ and $\Phi$ 
represent the gauge field strengths and chiral superfields of the 
SSM.  Interactions of the form $\int\! d^4\theta\, X_i^\dagger X_i 
\Phi^\dagger \Phi/\Lambda_i^2$ do not provide necessary $R$-breaking 
transmission, unless $X_i$ has a lowest component vev giving effectively 
$A$-term operators.  For the remainder of this section, we will absorb 
any vev for $X_i$ into the coefficients of the corresponding operators. 
Note that $R$-preserving operators can still play an important role 
in generating relevant effects when combined with operators that do 
transmit $R$-breaking.

\begin{figure}
\begin{center}
\includegraphics[scale=0.55]{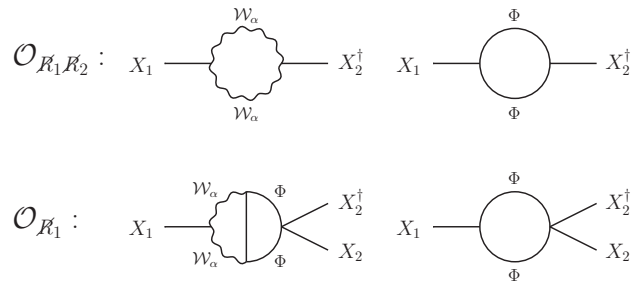}
\end{center}
\caption{Feynman diagrams which induce direct couplings between 
 sectors 1 and 2.  There is always at least one factor of $1/16\pi^2$ 
 coming from a loop of SSM fields.  Depending on the details of the 
 underlying theory, there may be additional loop factors, for instance 
 if the $\int\! d^2\theta\, X_{1} {\cal W}^\alpha {\cal W}_\alpha/\Lambda_1$ 
 coupling itself is generated at one loop.}
\label{fig:diagrams}
\end{figure}
We can characterize the induced couplings between sectors 1 and 2 
according to whether they violate $U(1)_{R_1}$, $U(1)_{R_2}$, or both. 
These couplings are generated by the diagrams in Fig.~\ref{fig:diagrams}, 
and have the form (after absorbing any vev for $X_i$ into the operator 
coefficients)
\bea
  {\cal O}_{\sslash{R}_1 \sslash{R}_2} 
  &\approx& \left(\frac{1}{16\pi^2}\right)^{n_{12}} 
    \int\! d^4\theta\, X_1 X_2^\dagger + {\rm h.c.},
\nonumber\\
  {\cal O}_{\sslash{R}_1} 
  &\approx& \left(\frac{1}{16\pi^2}\right)^{n_1} 
    \frac{1}{{\rm max}\{\Lambda_1, \Lambda_2\}} 
    \int\! d^4\theta\, X_1 X_2^\dagger X_2 + {\rm h.c.},
\nonumber\\
 {\cal O}_{\sslash{R}_2} 
  &\approx& \left(\frac{1}{16\pi^2}\right)^{n_2} 
    \frac{1}{{\rm max}\{\Lambda_1, \Lambda_2\}} 
    \int\! d^4\theta\, X_2 X_1^\dagger X_1 + {\rm h.c.},
\nonumber\\
\label{eq:direct}
\eea
where we have included the coefficients in the Lagrangian terms in the 
definitions of ${\cal O}$'s.  We now consider each of these operators 
in turn.

\subsection{Effects on Goldstini}
\label{subsec:effect-G}

If both $U(1)_{R_1}$- and $U(1)_{R_2}$-breaking effects 
exist and are transmitted, then the kinetic mixing operator 
${\cal O}_{\sslash{R}_1 \sslash{R}_2}$ will arise.  Note that 
$n_{12} \geq 1$, since there is always at least one loop of SSM 
fields involved in the diagram (if the gaugino mass operators 
themselves are generated at one loop, for instance as in gauge 
mediation, then $n_{12} = 3$).%
\footnote{The loop factor may not exist if the SSM sector contains 
 a singlet that directly mixes with SUSY breaking fields.  We assume 
 that such a singlet does not exist.}
However, since this operator is separately holomorphic in sector~$1$ 
and sector~$2$ fields, it separately preserves SUSY$_1$ and SUSY$_2$ 
in the limit in which derivatively coupled terms are neglected---hence, 
this operator does not contribute to $m_a$.%
\footnote{Incidentally, the same argument also shows that any operators 
 of the form $\int\! d^4\theta\, f_1(X_1) f_2(X_2^\dagger)$ do not 
 contribute to $m_a$.}
The only effect of ${\cal O}_{\sslash{R}_1 \sslash{R}_2}$ is to modify 
the kinetic term of $\zeta$ by an order $(1/16\pi^2)^{n_{12}}$ fraction, 
inducing $\delta m_a/m_a$ of the same size.  If there are more than two 
sectors which couple to the SSM in this way, then kinetic mixings of 
this order will be generated among all the $\zeta_a$.

If only $U(1)_{R_1}$-breaking effects are transmitted, then 
${\cal O}_{\sslash{R}_1}$ is generated, where again $n_1 \geq 1$ 
because there is at least one loop of SSM fields.  This operator 
yields a contribution to the goldstini masses
\bea
  {\cal O}_{\sslash{R}_1} 
  &\supset& \frac{(1/16\pi^2)^{n_1} F_2}{2\,{\rm max}\{\Lambda_1, \Lambda_2\}} 
    \left( \frac{F_2}{F_1} \eta_1^2 + \frac{F_1}{F_2} \eta_2^2 
    - 2 \eta_1 \eta_2 \right)
\nonumber\\
  &\rightarrow& \frac{1}{2} \left(\frac{1}{16\pi^2}\right)^{n_1}\!\! 
    \frac{\Feff}{\cos\theta\, {\rm max}\{\Lambda_1, \Lambda_2\}}\, \zeta^2,
\label{eq:zeta-2}
\eea
where in the last equation we have assumed that the only sectors breaking 
SUSY are sectors 1 and 2, and have plugged in for the mixing angles in 
\Eq{eq:goldstinibasis}.  Obviously, an identical analysis can be performed 
when only $U(1)_{R_2}$ is broken.

If neither of $U(1)_{R_1}$ or $U(1)_{R_2}$ breaking is transmitted, 
the goldstini Majorana masses cannot deviate from $2 m_{3/2}$.  The 
goldstini, however, may still obtain Dirac masses with fermions of 
$R$-charge $-1$.  For instance, consider $\int\! d^4\theta\, X_1^\dagger 
X_1 S^\dagger_2 S_2$, which is an $R$-symmetric coupling between a SUSY 
breaking field in sector~$1$ and a spectator field in sector~$2$ which 
does not have an $F$-component vev.  For $\langle S_2 \rangle \neq 0$, 
this operator induces a Dirac mass between the goldstino, $\zeta$, in 
$X_1$ and the fermionic component of $S_2$.  The effect from this class 
of operators, however, is generically smaller than that expected from 
${\cal O}_{\sslash{R}_1}$ and ${\cal O}_{\sslash{R}_2}$ for natural 
values of $\langle S_2 \rangle \sim O(\sqrt{F_2})$.

The operators ${\cal O}_{\sslash{R}_1}$ and ${\cal O}_{\sslash{R}_2}$ 
can potentially produce large corrections to the goldstini masses. 
However, since they are suppressed by ${\rm max}\{\Lambda_1, \Lambda_2\}$, 
we find that in most cases these corrections are
\be
  \delta m_a \simlt \left(\frac{1}{16\pi^2}\right)^n \tilde{m},
\ee
where $n \geq 1$ and $\tilde{m}$ is the scale for the SSM superpartner 
masses, which we have taken to be common for the gauginos and scalars. 
Therefore, if the gravitino mass is not substantially smaller than the 
superpartner masses, as in the case where $\Lambda_{1,2}$ are taken 
near the gravitational scale, then the relation $m_a = 2m_{3/2}$ will 
receive only small corrections.  The situation is model dependent if 
the gravitino is much lighter.  The corrected goldstini masses, however, 
are still significantly smaller than $\tilde{m}$, so that the SSM 
superpartners can decay into them.

The matrix $V_{ia}$, defined by \Eq{eq:V_ia}, is determined to diagonalize 
the goldstini mass matrix
\be
  {\cal L} = -\frac{1}{2} m_{ij} \hat{\eta}_i \hat{\eta}_j + {\rm h.c.}
\,\,\rightarrow\,\,
  -\frac{1}{2} m_a \zeta_a^2 + {\rm h.c.},
\ee
where $\hat{\eta}_i = \eta_i - (F_i/\Feff) \eta_{\rm long}$ is the 
goldstini field with the eaten mode projected out, and
\be
  m_{ij} = 2 m_{3/2}\, \delta_{ij} + \delta m_{ij},
\label{eq:mass-mat}
\ee
with $\delta m_{ij}$ representing the effects from the operators in 
\Eq{eq:direct}.  At the zero-th order in $\delta m_{ij}/m_{3/2}$ expansion, 
$V_{ia}$ is the $N \times (N-1)$ part of an orthogonal matrix preserving 
the first term of \Eq{eq:mass-mat}.  Since the angles of this matrix 
are determined by a perturbation, $\delta m_{ij}$, on the unit matrix 
$2 m_{3/2} \delta_{ij}$, they are typically of order unity.

Finally, we note that none of the operators discussed above affects the 
mass of the eaten mode, $\eta_{\rm long}$.  This is consistent with the 
general argument in \Sec{subsec:2-stuckelberg}.

\subsection{Other Corrections}
\label{subsec:other}

We have seen that the corrections to the goldstini masses from induced 
interactions between SUSY breaking sectors are generically small.  If 
there are tree-level direct couplings between these sectors, their effects 
can be studied similarly, following the analysis above.  The goldstini 
masses are also corrected if there is a deviation from the assumption 
that SUSY is broken in the global limit.  This effect is discussed 
briefly in the Appendix.

At loop level, the goldstini masses receive corrections from anomaly 
mediated effects, which exist even in the sequestered limit.  Using the 
non-linear parameterization, we can calculate the corrections and find
\be
  \delta m_{ij} = -\gamma_i m_{3/2}\, \delta_{ij},
\label{eq:anom}
\ee
where $\gamma_i$ is the anomalous dimension of $X_i$ defined by 
$d\ln Z_{X_i}/d\ln\mu_R = -2 \gamma_i$.%
\footnote{If the $X_i$ vev is nonzero in the basis where the $X_i$ 
 linear term vanishes in the K\"{a}hler potential, there is an additional 
 contribution $\delta m_{ij} = \dot{\gamma}_i x_i^* m_{3/2}^2/2 F_i$, 
 where $\dot{\gamma_i} = d\gamma_i/d\ln\mu_R$.  This contribution is 
 generically much smaller than that in \Eq{eq:anom}.}
Naturally, these contributions are loop suppressed.%
\footnote{Corrections of similar size may also be induced by direct 
 couplings between the SSM and SUSY breaking sectors.  For example, 
 loops of SSM states may generate holomorphic operators like 
 $\int\! d^4\theta\, X_1 X_2$, giving corrections loop suppressed 
 compared with $m_{3/2}$.}
Note that the eaten mode, $\eta_{\rm long}$, does not receive such 
a correction.

\section{Interactions with the SSM Sector}
\label{sec:vis-int}

In this section, we show how the goldstini couple to the SSM.  As per 
usual, the gravitino couples to the SSM fields through its eaten goldstino 
component, $\eta_{\rm long}$, whose interactions to a chiral multiplet 
take the form
\be
  {\cal L}_{\rm int} 
  \supset \frac{1}{\Feff} \Bigl( \sum_i \tilde{m}_i^2 \Bigr)
    \eta_{\rm long} \psi \phi^\dagger,
\ee
where $\psi$ and $\phi$ are the fermionic and bosonic components of 
a chiral superfield $\Phi$ of the SSM, and $\tilde{m}_i$ is the soft 
mass contribution to this field from sector~$i$.  The interactions 
to a vector multiplet are given by
\be
  {\cal L}_{\rm int} 
  \supset -\frac{i}{\sqrt{2}\Feff} \Bigl( \sum_i \tilde{m}_i \Bigr) 
    \eta_{\rm long} \sigma^{\mu\nu} \lambda F_{\mu\nu},
\ee
where $\lambda$ is the gaugino, and $\tilde{m}_i$ is the contribution 
to its mass from sector~$i$.

The couplings of the uneaten goldstini to the SSM fields are different 
from those of the gravitino.  We first consider those to chiral multiplets. 
The couplings of $\eta_i$ to the SSM states can be obtained by using 
\Eq{eq:nonlinparam} in
\be
  {\cal L} = \sum_i \frac{1}{\Lambda_i^2} \int\! d^4\theta\, 
    X_i^\dagger X_i \Phi^\dagger \Phi,
\ee
giving scalar mass contributions $\tilde{m}_i^2 = -F_i^2/\Lambda_i^2$. 
The interactions of the uneaten goldstini are then
\be
  {\cal L}_{\rm int} 
  \supset \frac{1}{\Feff} \sum_{i,a} 
    \frac{\tilde{m}_i^2 V_{ia}}{r_i} \zeta_a \psi \phi^\dagger,
\ee
where $F_i \equiv r_i \Feff$ ($\sum_i r_i^2 = 1$), and we have used 
\Eq{eq:V_ia}.  In the case where there are only two SUSY breaking 
sectors, these interactions become
\bea
  {\cal L}_{\rm int} 
  &\supset& -\frac{1}{\Feff} 
    (\tan\theta\, \tilde{m}_1^2 - \cot\theta\, \tilde{m}_2^2) 
    \zeta \psi \phi^\dagger
\nonumber\\
  &\approx& -\left( \frac{F_2}{F_1^2} \tilde{m}_1^2 
    - \frac{1}{F_2} \tilde{m}_2^2 \right) \zeta \psi \phi^\dagger + \ldots,
\label{eq:scalarint}
\eea
where in the last equation we have assumed $F_1 \gg F_2$ and approximated 
$\Feff$ by $F_1$.

\begin{figure}
\begin{center}
\includegraphics[scale=0.9]{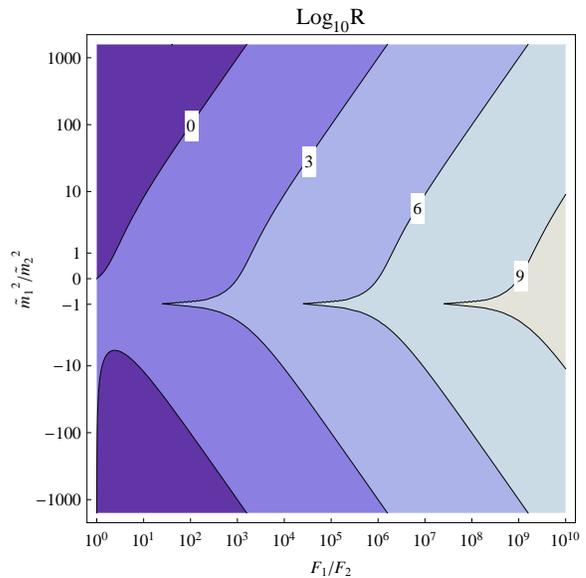}
\end{center}
\caption{A contour plot of $R$ in \Eq{eq:Rcoeff} as a function of 
 $\tilde{m}_1^2/\tilde{m}_2^2$ and $F_1/F_2$.  When $|\tilde{m}_1^2| 
 \simlt |\tilde{m}_2^2|$, $R$ is greater than unity for a wide range 
 of $F_1/F_2$, so that the SSM sector fields couple more strongly 
 to the uneaten goldstino than to the gravitino.}
\label{fig:Rplot}
\end{figure}

In the two sector case, it is useful to define the quantity
\be
  R = \left| \frac{\textrm{coefficient of } \zeta \psi \phi^\dagger}
    {\textrm{coefficient of } \eta_{\rm long} \psi \phi^\dagger} \right|,
\label{eq:Rcoeff}
\ee
which characterizes the relative interaction strength of the SSM sector fields 
to the uneaten goldstino versus the gravitino.  In Fig.~\ref{fig:Rplot}, 
we plot $R$ as a function of $\tilde{m}_1^2/\tilde{m}_2^2$ and $F_1/F_2 = 
\cot\theta$.  We find that $R > 1$ more or less whenever $|\tilde{m}_1^2| 
\simlt |\tilde{m}_2^2|$---the SSM fields generically couple more strongly 
to the uneaten goldstino in this case.

The couplings of the goldstini to vector multiplets can be worked out 
similarly, and are given by
\be
{\cal L}_{\rm int}
  \supset -\frac{i}{\sqrt{2}\Feff} \sum_{i,a} 
    \frac{\tilde{m}_i V_{ia}}{r_i} 
    \zeta_a \sigma^{\mu\nu} \lambda F_{\mu\nu}.
\ee
If there are only two SUSY breaking sectors,
\bea
{\cal L}_{\rm int}
  &\supset& \frac{i}{\sqrt{2}\Feff} 
    (\tan\theta\, \tilde{m}_1 - \cot\theta\, \tilde{m}_2) 
    \zeta \sigma^{\mu\nu} \lambda F_{\mu\nu}
\nonumber\\
  &\approx& \frac{i}{\sqrt{2}} 
    \left( \frac{F_2}{F_1^2} \tilde{m}_1 - \frac{1}{F_2} \tilde{m}_2 \right) 
    \zeta \sigma^{\mu\nu} \lambda F_{\mu\nu} + \ldots,
\label{eq:gauginoint}
\eea
where we have set $\Feff \approx F_1$ in the last line.  As in the case of 
chiral multiplets, the couplings to the uneaten goldstino are generically 
stronger than those to the gravitino for $|\tilde{m}_1| \simlt |\tilde{m}_2|$.

\section{Collider Phenomenology}
\label{sec:collider}

Goldstini may be probed directly or indirectly at the LHC.  In what 
follows, we consider a minimal setup in which SUSY is broken in two 
separate sectors, yielding a gravitino $\tilde{G}$ and a single uneaten 
goldstino $\zeta$.  This scenario preserves most of the salient features 
of our general framework.

We focus our analysis on the regime in which $|\tilde{m}_1| \simlt 
|\tilde{m}_2|$, so that the SSM fields couple more strongly to $\zeta$ 
than to $\tilde{G}$.  This includes the case from Fig.~\ref{fig:setup-1} 
where a conventional SUSY breaking scenario is augmented by an additional, 
completely sequestered SUSY breaking sector with a higher SUSY breaking 
scale.  Below we explore five classes of novel LHC signatures which can 
occur within our framework.  We assume $R$-parity conservation throughout.

\subsection{``Gravitino'' with a Wrong Mass-Interaction Relation}
\label{subsec:wrong-rel}

Suppose that sector~$2$ which has $F_2$ ($\ll F_1$) gives masses to 
all the SSM superpartners.  In this case, $\zeta$ couples more or 
less universally to all the SSM states, so that $\zeta$ looks like 
the ``gravitino'' when interpreted in the conventional framework. 
This apparent ``gravitino'', however, has a wrong mass-interaction 
relation.  Indeed, its interactions are controlled by $F_2$ 
(cf.~\Eqs{eq:scalarint}{eq:gauginoint} when $|\tilde{m}_1| \simlt 
|\tilde{m}_2|$), but its mass is controlled by $F_1$ (since $m_\zeta 
\simeq 2 F_1/\sqrt{3}\MPl$).  This is different from the true 
gravitino, whose interactions and mass are controlled by a single 
parameter $\Feff$.  Said another way, the goldstino has 
a mass which is a factor of $\simeq 2 F_1/F_2$ larger than that 
of a conventional gravitino with a comparable interaction strength. 

Suppose that $\zeta$ (and $\tilde G$) is lighter than all of the 
SSM superpartners, which we assume throughout this subsection.  In 
this case, all the SUSY cascade will terminate with the lightest 
observable-sector supersymmetric particle (LOSP) decaying dominantly 
into $\zeta$.%
\footnote{The goldstino $\zeta$ will decay further into the 
 gravitino through intermediate SSM states.  As we will see in 
 \Sec{subsec:cosmo-stable}, this decay is very slow, so that 
 $\zeta$ can be regarded as a stable particle.}
As in conventional gauge mediation, if $\sqrt{F_2} \simlt 
10^7~{\rm GeV}$ this decay may occur inside the detector; in 
particular, for small $\sqrt{F_2} \sim O(10~\mbox{--}~100~{\rm TeV})$ 
it is prompt.  Such a decay can provide a distinct signature at the 
LHC~\cite{Dimopoulos:1996vz}.  A unique aspect in our framework is 
that the mass of the escaping state can be significant, e.g.\ $\simgt 
O(10~{\rm GeV})$ for $\sqrt{F_1} \approx O(10^9~\mbox{--}~10^{10}~{\rm 
GeV})$, which cannot be the case in conventional gauge mediation. 
Therefore, if we can somehow measure a nonzero mass of this state, 
perhaps using methods similar to those discussed in Ref.~\cite{Cheng:2007xv}, 
we can discriminate the present scenario from the usual one.  These 
signals will be especially distinct if the LOSP is the bino (yielding 
two photons in the final state) or if a charged slepton LOSP decay 
leaves a displaced kink in the tracking detector.  For massive 
escaping particles, such signals are hardly obtained in the 
conventional framework.%
\footnote{The signals cannot be mimicked by a LOSP decay into the QCD 
 axino either, since given an axion decay constant avoiding laboratory and 
 astrophysical bounds, the decay occurs always outside the detector.}

If the LOSP is charged, then there can be a striking signature arising 
from a long-lived charged state.  For $\sqrt{F_2} \simgt 10^6~{\rm GeV}$, 
the LOSP may still live long enough that its mass and lifetime can be 
precisely determined by, e.g., velocity measurements and by observing 
decays of stopped LOSPs either inside a main detector~\cite{CMS-stopped} 
or in a proposed stopper detector~\cite{Hamaguchi:2004df}.  Measurement 
of LOSP decays also allows us to determine the mass of the invisible state 
to which the LOSP decays, as long as it is larger than $O(10~{\rm GeV})$. 
In fact, the charged LOSP arises naturally in many theoretical constructions. 
For example, the right-handed stau can easily be the LOSP if SUSY breaking 
is transmitted from sector~$2$ to the SSM sector via gauge or gaugino 
mediation.  The LOSP may also be a selectron or smuon if there is 
a controlled source of flavor violation, which leads to a spectacular 
signal of monochromatic electrons or muons~\cite{Nomura:2007ap}.

In the conventional scenario, the charged LOSP decays into 
the gravitino.  Since the lifetime of the LOSP and the gravitino 
mass are related by $\Feff$, one can indirectly measure the Planck 
scale~\cite{Buchmuller:2004rq}
\bea
  &\Gamma_{\tilde{l} \rightarrow l \tilde{G}} 
  \simeq \frac{m_{\tilde{l}}^5}{16\pi \Feff^2},
\qquad
  m_{3/2} \simeq \frac{\Feff}{\sqrt{3} \MPl}&
\nonumber\\
  &\Longrightarrow \qquad
  \MPl^2 \simeq \frac{m_{\tilde{l}}^5}
    {48\pi \Gamma_{\tilde{l} \rightarrow l \tilde{G}} m_{3/2}^2},&
\eea
where we have adopted notation appropriate for a slepton LOSP.  However, 
this is not the case if the LOSP instead decays into the uneaten goldstino 
$\zeta$, since the goldstino mass and decay constant are controlled 
by separate parameters and thus a priori unrelated.  Specifically, for 
$F_1 \gg F_2$, we will mismeasure $\MPl$ by a factor of $F_2/2 F_1$ if 
we misinterpret $\zeta$ as a conventional gravitino
\bea
  &\Gamma_{\tilde{l} \rightarrow l \zeta} 
  \simeq \frac{m_{\tilde{l}}^5}{16\pi F_2^2},
\qquad
  m_\zeta \simeq \frac{2 F_1}{\sqrt{3} \MPl}&
\nonumber\\
  &\Longrightarrow \qquad
  \MPl^2 \simeq \left( \frac{2 F_1}{F_2} \right)^2 \frac{m_{\tilde{l}}^5}
    {48\pi \Gamma_{\tilde{l} \rightarrow l \zeta} m_\zeta^2},&
\eea
which would reveal that the particle to which the LOSP is decaying is not 
the gravitino.%
\footnote{The LOSP decay product, however, may be the QCD axino $\tilde{a}$. 
 Discriminating between $\zeta$ and $\tilde{a}$ using the lifetime 
 measurement will be difficult because values of $\Gamma_{\tilde{l} 
 \rightarrow l \zeta}$ and $\Gamma_{\tilde{l} \rightarrow l \tilde{a}}$ 
 mostly overlap in relevant parameter regions, especially if we allow 
 the axion decay constant to be in the so-called anthropic range. 
 The discrimination, however, may be possible by studying detailed 
 structures of radiative three-body decays~\cite{Brandenburg:2005he}.}

\subsection{Gravitinoless Gauge Mediation}
\label{subsec:pheno-GMSB}

Thus far we have considered a case where $\zeta$ and $\tilde{G}$ 
are lighter than the LOSP.  However, since the masses of $\zeta$ and 
$\tilde{G}$ are both controlled by the largest SUSY breaking scale $F_1$, 
these states can be heavier than all the SSM superpartners even if $F_2$ 
(and the corresponding mediation scale $\Lambda_2$) is small.  As a 
consequence, the LOSP may be stable even if SSM superpartners obtain 
their masses primarily from a sector having low SUSY breaking and 
mediation scales.

This allows for a canonical gauge mediation spectrum without a light 
gravitino, and hence with neutralino dark matter.  A scenario with 
similar phenomenology was considered before in Ref.~\cite{Nomura:2001ub}. 
In our context, it arises as a special case of the general framework 
of multiple SUSY breaking.

\subsection{Measuring the ``Two''}
\label{subsec:pheno-two}

We have seen that the uneaten goldstino $\zeta$ may appear as a 
``gravitino'' with a wrong mass-interaction relation, or may be heavier 
than the LOSP, making it irrelevant for collider experiments.  Is there 
a situation in which we might directly observe both $\zeta$ and $\tilde{G}$ 
and measure their detailed properties, in particular their mass ratio? 
The answer to this question is yes.

\begin{figure}
\begin{center}
\includegraphics[scale=.9]{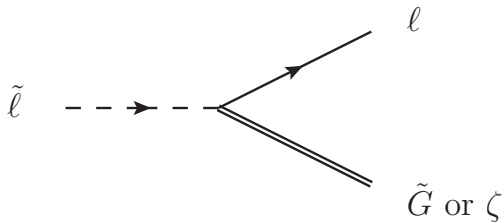}
\end{center}
\caption{If $F_1 \approx F_2$ and $\tilde{m}_1 \approx \tilde{m}_2$, then 
 the SSM states couple to $\zeta$ and $\tilde{G}$ with similar strengths. 
 In particular, if $\zeta$ and $\tilde{G}$ are lighter than all the 
 SSM superpartners, then the LOSP decays into $\zeta$ or $\tilde{G}$ 
 with non-negligible branching ratios.  This allows for the possibility 
 of measuring the masses of both $\zeta$ and $\tilde{G}$, providing 
 smoking gun evidence for multiple sector SUSY breaking.}
\label{fig:slepton}
\end{figure}

Suppose that two SUSY breaking sectors have comparable SUSY breaking 
strengths, $F_1 \approx F_2$, and contribute comparably to the masses 
of SSM superpartners, $\tilde{m}_1 \approx \tilde{m}_2$.%
\footnote{Such a situation may naturally be realized if environmental 
 selection acts on superpartner masses through the requirement on the 
 weak scale, and the two SUSY breaking sectors have comparable mediation 
 scales, e.g., around the string scale.}
In this case, $\zeta$ and $\tilde{G}$ couple to SSM states with similar 
strengths.  Therefore, if both $\zeta$ and $\tilde{G}$ are lighter 
than all the SSM states, then the branching ratios of the LOSP to 
$\zeta$ and $\tilde{G}$ are both non-negligible, as illustrated in 
Fig.~\ref{fig:slepton} for the case of the slepton LOSP.

If $m_\zeta, m_{3/2} \simgt O(10~{\rm GeV})$, these masses can be determined 
by measuring the decays of long-lived charged LOSPs, using the same 
techniques as in \Sec{subsec:wrong-rel}.  This mass range corresponds to 
$\sqrt{F_1} \approx \sqrt{F_2} \approx O(10^9~\mbox{--}~10^{10}~{\rm GeV})$, 
so that the LOSP is long lived.  In the case that direct interactions 
between SUSY breaking sectors are small, this measurement will find two 
invisible states $X_{1,2}$ whose masses satisfy
\be
  m_{X_1}/m_{X_2} \approx 2.
\ee
This would be an unmistakable signature of the uneaten goldstino $\zeta$ 
(or goldstini $\zeta_a$ with a degenerate mass), and hence smoking gun 
evidence for multiple sector SUSY breaking.

\subsection{Difermions with Fixed Ratios}
\label{subsec:pheno-4fermions}

Distinct signatures may also arise if sectors 1 and 2 couple to the SSM 
in a more elaborate fashion.  In particular, if one of these sectors 
preserves an (approximate) $R$ symmetry, then the SSM gaugino masses 
are entirely generated by the other sector.  This will affect the couplings 
of the SSM states to $\zeta$, and can substantially change phenomenology.

\begin{figure}
\begin{center}
\includegraphics[scale=.9]{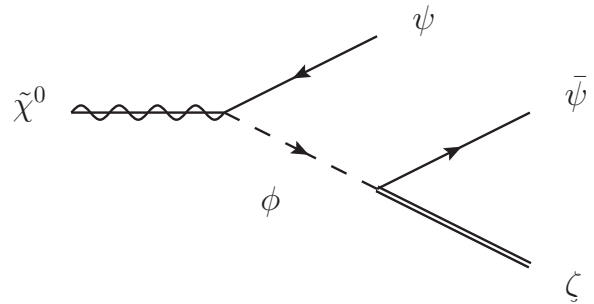}
\end{center}
\caption{If $F_1 \gg F_2$ and the SSM gaugino masses arise from sector~$1$ 
 alone, then a bino-like LOSP can decay into $\zeta$ and two standard 
 model fermions $\psi \bar{\psi}$ through an off-shell scalar $\phi$, 
 which is the superpartner of $\psi$.  For $m_{\tilde{\chi}^0}^2 \ll 
 m_\phi^2$ and $\tilde{m}_1^2 \ll \tilde{m}_2^2$, the branching fraction 
 into each $\psi \bar{\psi}$ is entirely determined by the hypercharge 
 of this field.}
\label{fig:neutralino}
\end{figure}
Consider a situation that the two sectors have $F_1 \gg F_2$ and contribute 
comparably to the scalar masses, but that the gaugino masses arise solely 
from sector~$1$.  This is true if sector~$2$ preserves an $R$ symmetry. 
In this setup, the SSM scalars couple strongly to $\zeta$, while the gauginos 
do so only very weakly.  Therefore, if the LOSP is a bino-like neutralino, 
it decays either via $\tilde{\chi}^0 \rightarrow Z \zeta, h \zeta$ through 
its Higgsino fraction, or via $\tilde{\chi}^0 \rightarrow \zeta \psi 
\bar{\psi}$ through the off-shell SSM scalar $\phi$ which is the superpartner 
of a standard model fermion $\psi$ (see Fig.~\ref{fig:neutralino}).  If 
$\tilde{\chi}^0$ has a significant Higgsino fraction, $\simgt O(0.1)$, 
and its decay into $Z$ or $h$ is not kinematically suppressed, then the 
former modes dominate.  In this case the signature would look like the 
Higgsino LOSP decaying into $\zeta$, even if the LOSP is bino-like.

If the above conditions are not met, the three-body decay $\tilde{\chi}^0 
\rightarrow \zeta \psi \bar{\psi}$ dominates.  In the limit that 
$m_{\tilde{\chi}^0}^2 \ll m_\phi^2$, the amplitude of this decay is 
proportional to $Y \tilde{m}_2^2/(\tilde{m}_1^2 + \tilde{m}_2^2)$, where 
$Y$ is the hypercharge of $\psi/\phi$ and $\tilde{m}_{1,2}^2$ are the 
contributions to the $\phi$ mass-squared from each sector.  Interestingly, 
for $\tilde{m}_1^2 \ll \tilde{m}_2^2$, the dependence on the $\phi$ mass 
drops out completely due to a cancellation between the propagator and the 
vertex factor.  Therefore, in this parameter region, the ratios to various 
final states $\psi \bar{\psi}$ are entirely fixed by $Y$, giving
\be
  q\bar{q} : b\bar{b} : t\bar{t}: e\bar{e}: \mu\bar{\mu} : \tau\bar{\tau} 
  \simeq 44 : 5 : 17 : 15 : 15 : 15,
\ee
where $q = u,d,s,c$.  (There is also a completely invisible mode to 
neutrinos, and the $t\bar{t}$ mode may have a kinematic suppression. 
If $m_{\tilde{\chi}^0} > m_\zeta + 2 m_h$, then $\tilde{\chi}^0 \rightarrow 
\zeta h h$ is also possible, whose rate depends on the masses of the 
Higgs/Higgsino.)  This provides a unique signature of the setup considered 
here.  Note that the decay of $\tilde{\chi}^0$ may also occur with 
a displaced vertex, since the $\tilde{\chi}^0$ lifetime can be long 
in some regions of parameter space.

\subsection{Displaced Monojets}
\label{subsec:pheno-jets}

Another spectacular signal may arise if the SSM scalars are much 
heavier than the gauginos, as in split SUSY~\cite{ArkaniHamed:2004fb}. 
In particular, suppose that sector~$2$ provides weak scale masses to 
all of the SSM superpartners, while sector~$1$ does so only for the 
scalars---this can easily occur if sector~$1$ preserves an $R$ symmetry. 
We also assume that the scalar masses from sector~$1$ are much greater 
than the weak scale.

\begin{figure}
\begin{center}
\includegraphics[scale=.75]{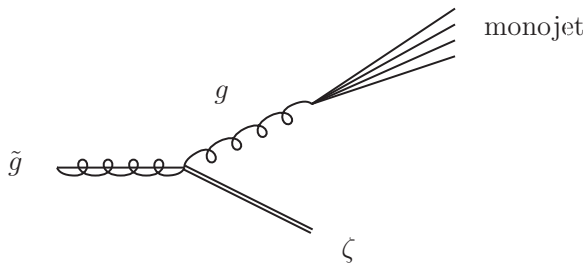}
\end{center}
\caption{If the squarks are sufficiently heavy, then the dominant gluino 
 decay channel may be $\tilde{g} \rightarrow g \zeta$, which appears 
 as a displaced gluino decaying into a monojet recoiling off of missing 
 energy.  If $\zeta$ is not at the bottom of the SUSY spectrum, then 
 the decay of the $\zeta$ will produce a second displaced vertex.}
\label{fig:monojet}
\end{figure}

If $m_\zeta < m_{\tilde{g}}$ and the squark masses are sufficiently 
large, $m_{\tilde{q}}^2 \simgt F_2/4\pi$, then the gluino prefers to 
decay directly into $\zeta$ and a gluon instead of cascade decaying 
through an off-shell squark.  While $\tilde{g} \rightarrow g \zeta$ 
will generically be slow, for $\sqrt{F_2} \simlt 10^7~{\rm GeV}$ it may 
occur within the detector.  This gives a distinct signal of a displaced 
gluino decaying into a monojet recoiling off of missing energy (see 
Fig.~\ref{fig:monojet}).

Furthermore, if $\zeta$ is not at the very bottom of the superpartner 
spectrum, it will further decay into lighter SSM states.  If the initial 
gluino decay occurs within the detector, then the $\zeta$ decay will 
also likely occur within the detector.  This provides a spectacular 
signature of a secondary displaced vertex corresponding to the decay 
of the uneaten goldstino $\zeta$.%
\footnote{While the signal of displaced monojets may be mimicked by 
 conventional gauge mediation models with the gluino LOSP, the signal 
 of a secondary displaced vertex cannot.}

\section{Cosmology}
\label{sec:cosmology}

As one might expect, goldstini cosmology is not very dissimilar from 
gravitino cosmology.  However, there are important differences arising 
from the fact that, unlike the gravitino, the goldstini have masses 
and couplings which are parametrically unrelated.  This affects cosmology 
especially when these fields are lighter than the LOSP, which we will 
focus in this section.

As with the collider signatures in the previous section, we focus 
on the case of two SUSY breaking sectors with $|\tilde{m}_1| \simlt 
|\tilde{m}_2|$.  We also assume that deviations from the sequestered 
limit are small: the uneaten goldstino $\zeta$ has a mass $m_\zeta 
\simeq 2 F_1/\sqrt{3} M_{\rm Pl}$ and couplings to SSM fields 
proportional to $1/F_2$.

We assume ``standard'' cosmological history throughout this section. 
Many of the constraints discussed below can be avoided if we deviate 
from this assumption, e.g., if there is late time entropy production 
at temperature significantly below the weak scale.

\subsection{Goldstini are Cosmologically Stable}
\label{subsec:cosmo-stable}

If the goldstino $\zeta$ is lighter than the LOSP, it decays into the 
gravitino via $\zeta \rightarrow \tilde{G} \psi \bar{\psi}$, where 
$\psi$ is a standard model fermion (arguments similar to the ones below 
will also hold for decays into photons).  As we will see, this is 
longer than the age of the universe, so we can treat both goldstino 
and gravitino as stable particles.

In the conventional SUSY picture, low energy theorems dictate that 
the contact interaction $\tilde{G} \tilde{G} \psi \bar{\psi}$ is 
controlled by $E^4/\Feff^2$, where $E$ is a typical energy scale of the 
reaction~\cite{Brignole:1996fn}.  While a complete description of goldstini 
low energy ``theorems'' is beyond the scope of this work, we note that 
$\zeta \tilde{G} \psi \bar{\psi}$ also scales like $E^4/\Feff^2$, albeit 
with a prefactor that depends on $\tilde{m}_i$ and $F_i$.  Consequently, 
the width of the goldstino is given parametrically by
\be
  \Gamma_{\zeta \rightarrow \tilde{G} \psi \bar{\psi}} 
  \approx \frac{1}{128\pi^3} \frac{m_\zeta^9}{\Feff^4} 
    \left(\frac{F_1}{F_2} \frac{\tilde{m}_2^2}{\tilde{m}_1^2+\tilde{m}_2^2} 
    \right)^2.
\ee
The shortest reasonable lifetime is then
\be
  \tau_{\zeta \rightarrow \tilde{G} \psi \bar{\psi}} 
  \approx 10^{22}~{\rm sec} 
    \left( \frac{\sqrt{F_2}}{100~{\rm TeV}} \right)^4 
    \left( \frac{100~{\rm GeV}}{m_\zeta} \right)^7,
\ee
so the goldstino is cosmologically stable.  In theories of multiple 
sector SUSY breaking, decay transitions among the goldstini will take 
even longer, since they are nearly degenerate in mass.

\subsection{Late Decaying LOSP}
\label{subsec:cosmo-BBN}

Late decays of the LOSP to the goldstino will produce electromagnetic 
and/or hadronic fluxes which can alter the abundances of light 
elements and ruin the successful predictions of big bang nucleosynthesis 
(BBN)~\cite{Khlopov:1984pf}.  To safely evade such bounds, one either 
needs a small relic density of LOSPs, or the LOSP must have a lifetime 
shorter than $\sim 100~{\rm sec}$.

For a conventional gravitino, BBN typically imposes a severe constraint 
$m_{3/2} \simlt (10^{-2}~\mbox{--}~1)~{\rm GeV}$~\cite{Kawasaki:2004qu}, 
where the precise values depend on the identity, mass, and abundance 
of the LOSP.  In our case, however, the mass and coupling strengths of 
the uneaten goldstino are parametrically unrelated.  Thus, the LOSP 
decay rate to the goldstino is a factor of $(F_1/F_2)^2$ greater than 
what one would expect for a comparable mass gravitino.  Said another 
way, the goldstino behaves like a ``gravitino'' to which the LOSP decays 
faster than it should.  Note that the usual LOSP to gravitino decay 
is now irrelevant, since the LOSP will primarily decay into the goldstino. 
As a consequence, a goldstino (and gravitino) mass in the range of 
$(1~\mbox{--}~100)~{\rm GeV}$ is easily compatible with BBN constraints 
in wide regions of parameter space.

\subsection{Overproduction in the Early Universe}
\label{subsec:cosmo-overprod}

Another issue of a stable goldstino is that it may be overproduced 
in the early universe.  For a comparable mass, the goldstino couples 
more strongly to the SSM states than the gravitino.  This property 
has helped to avoid the BBN problem, as discussed above, but may hurt 
the overproduction problem.  (We will see a way to sidestep this 
conclusion in the next subsection.)

Suppose that sector~$2$ provides sizable contributions to all 
of the SSM superpartners.  The goldstino will then couple to the 
SSM much like a conventional gravitino.  As in usual gravitino 
cosmology~\cite{Pagels:1981ke}, the bound from overproduction is 
avoided for $m_\zeta \simlt 0.2~{\rm keV}$, since then the relic 
goldstino abundance from early thermal plasma is sufficiently small.%
\footnote{Structure formation, however, provides a stronger bound 
 of $m_\zeta \simlt O(10~{\rm eV})$ in this case~\cite{Viel:2005qj}.}
For larger goldstino masses, there are upper bounds on the reheating 
temperature $T_R$ in order for the relic goldstino not to overclose 
the universe.

It is relatively straightforward to translate the usual bounds for 
a gravitino, $\hat{T}_R^{\max}$, into corresponding bounds for an 
uneaten goldstino, $T_R^{\max}$.  Since $\zeta$ has interaction strengths 
controlled by $F_2$, its {\it number} density $n_\zeta$ is (approximately) 
the same as that one would have computed for a gravitino with $m_{3/2} 
= F_2/\sqrt{3}\MPl$.  The {\it energy} density $m_\zeta n_\zeta$, 
however, is larger than that of a gravitino with the same mass by 
$m_\zeta/(F_2/\sqrt{3}\MPl) = 2F_1/F_2$, implying
\be
  T_R^{\max} \left( m_\zeta, F_2 \right) = \frac{F_2}{2F_1} 
    \hat{T}_R^{\max} \left( m_{3/2} = \frac{F_2}{\sqrt{3}\MPl} \right).
\ee
Note that this expression is not valid if $T_R$ is sufficiently, 
typically $O(10)$, smaller than the superpartner mass scale, since then 
processes of goldstino generation are not active.  Using the result 
for the standard gravitino scenario~\cite{Moroi:1993mb}, we then find%
\footnote{This bound assumes a gluino mass of $1~{\rm TeV}$.  In general, 
 $T_R^{\max}$ scales as $m_{\tilde{g}}^{-2}$.}
\be
  T_R^{\max} \approx 
    100~{\rm GeV}\, \Biggl( \frac{1~{\rm GeV}}{m_\zeta} \Biggr) 
    \Biggl( \frac{\sqrt{F_2}}{10^8~{\rm GeV}} \Biggr)^4,
\label{eq:TR-bound}
\ee
for $T_R^{\max} \simgt O(100~{\rm GeV})$; for $T_R \simlt O(100~{\rm GeV})$, 
the bound disappears.  The bound of \Eq{eq:TR-bound} can also be written as 
$T_R^{\max}(m_\zeta, F_2) = (F_2/2 F_1)^2 \hat{T}_R^{\max} (m_{3/2}=m_\zeta)$, 
so for $F_2 \ll F_1$ the reheating bound for the uneaten goldstino is 
significantly stronger than that for a comparable mass gravitino.

\subsection{Goldstini Dark Matter}
\label{subsec:cosmo-DM}

As we have seen, the constraint from BBN is avoided if the LOSP 
lifetime is sufficiently short, corresponding to
\be
  \sqrt{F_2} \simlt (10^8~\mbox{--}~10^9)~{\rm GeV}.
\label{eq:F2-upper}
\ee
Then if $T_R$ saturates the bound of \Eq{eq:TR-bound}, $T_R \simeq 
T_R^{\max}$, the uneaten goldstino will comprise all of dark matter. 
(Here we have assumed that the $\zeta$ abundance generated by possible 
late LOSP decays is small.)  The required reheating temperature, however, 
is generically small in this case.

The strong bound of \Eq{eq:TR-bound} on the reheating temperature 
was obtained by assuming that $\zeta$ couples to all the SSM states 
with the strengths $\approx 1/F_2$.  However, this need not be the case. 
Consider, for example, that sectors 1 and 2 contribute comparably to 
the SSM scalar masses, but the gauginos obtain masses only from sector~$1$. 
This is the setup considered in \Sec{subsec:pheno-4fermions}, and occurs 
naturally if sector~$2$ preserves an $R$ symmetry.  In this case, $\zeta$ 
couples to the scalars with the strengths $\approx 1/F_2$, but to the 
gauginos with $\approx F_2/F_1^2$, which are much weaker for $F_2 \ll F_1$.

The absence of strong $\zeta$-gaugino interactions drastically changes 
the constraint from overproduction, since the standard reheating 
bound, \Eq{eq:TR-bound}, is dominated by $\zeta$ production from 
scattering involving the gluino.  In the absence of these interactions, 
the constraint comes from $\zeta$ production from early scalar scatterings 
and decays, which will be satisfied for
\be
  \sqrt{F_2} \simgt 10^8~{\rm GeV} 
    \left( \frac{m_\zeta}{1~{\rm GeV}} \right)^{1/4}.
\label{eq:F2-lower}
\ee
Therefore, if \Eqs{eq:F2-upper}{eq:F2-lower} are simultaneously satisfied, 
and if the LOSP is a scalar, then the constraints from both BBN and 
$\zeta$ overproduction can be avoided even for very large $T_R$.%
\footnote{If the LOSP is a gaugino, the dominant decay is the three-body 
 decay mode from \Sec{subsec:pheno-4fermions}, which faces more stringent 
 BBN constraints because of phase space suppression.}
Whether this is indeed possible, however, will require a more detailed 
analysis because of $O(1~\mbox{--}~10)$ uncertainties in our estimates 
of the constraints.

If $\sqrt{F_2}$ saturates \Eq{eq:F2-lower}, the generated $\zeta$ can 
comprise all of dark matter without any additional contributions.  Assuming 
that \Eq{eq:F2-upper} is satisfied, the bound on $T_R$ comes only from 
the usual gravitino overproduction, which is rather weak if $m_\zeta 
\simeq 2 m_{3/2}$ is not much smaller than the weak scale, e.g. if 
$\sqrt{F_1} \approx (10^9~\mbox{--}~10^{10})~{\rm GeV}$.  If $\sqrt{F_2}$ 
satisfies but does not saturate \Eq{eq:F2-lower}, then the $\zeta$ 
abundance must be dominated by late LOSP decays in order for $\zeta$ 
to be dark matter~\cite{Feng:2003xh}:
\be
  \Omega_\zeta \simeq \frac{m_\zeta}{m_{\rm LOSP}} \Omega_{\rm LOSP},
\ee
where $\Omega_{\rm LOSP}$ is the fractional contribution of the LOSP 
to the critical density if it did not decay into $\zeta$.  Since 
$\Omega_{\rm LOSP}$ is controlled by the standard WIMP parametrics, 
so is $\Omega_\zeta$ if $m_\zeta$ is not significantly below 
$m_{\rm LOSP}$.

\section{Discussion}
\label{sec:discuss}

The hypothesis of single sector SUSY breaking has by and large dictated 
the standard picture of SUSY phenomenology at colliders and in cosmology. 
In the conventional scenario, the (only) goldstino is eaten by the 
gravitino, whose mass and coupling strength to SSM fields are inextricably 
and sometimes problematically related.

Motivated by top-down considerations, we have relaxed this underlying 
assumption and considered the possibility that a multiplicity of 
sectors break SUSY, yielding a corresponding multiplicity of goldstini. 
Intriguingly, \emph{even when these additional sectors are completely 
sequestered from the SSM, this can have a drastic effect on LHC collider 
phenomenology}.  Ultimately this occurs because the gravitino eats 
a linear combination of the goldstini, and in a curious twist on the 
conventional narrative, what would have been our gravitino is replaced 
by a linear combination of the uneaten goldstini.

A key result of this paper is that all of the uneaten goldstini receive 
an irreducible and universal mass $m_a = 2 m_{3/2}$ from SUGRA effects, 
as long as SUSY is broken in the global limit.  As a consequence, the 
SSM fields can have sizable couplings to the goldstini, whose masses 
and decay constants are a priori unrelated.  This greatly expands the 
realm of phenomenological possibilities.  In particular, we considered 
a number of novel collider signatures, including anomalous neutralino 
and slepton decays, gravitinoless gauge mediated spectra, and monojet 
signals from (multiple) displaced vertices.

A true smoking gun signature of multiple sector SUSY breaking will 
exist if a charged LOSP has sizable branching ratios to both the 
gravitino and at least one goldstino.  In this case, the mass ratio 
between the gravitino and goldstino may be accurately measured in 
a stopper detector, and a ratio of $2$ would give dramatic evidence 
towards the scenario considered in this paper.

There are many possible directions for future work.  While we have 
concentrated on the scenario where each SUSY breaking sector is $F$-term 
dominated, there is of course the possibility that one or more sectors 
experience $D$-term or ``almost no-scale'' SUSY breaking.  In the 
latter case, there is significant mixing between gravitational modes 
and SUSY breaking fields, and as previewed in the Appendix, the goldstini 
masses can deviate significantly from $2 m_{3/2}$.  Moreover, while most 
of the phenomenological analyses in this work have focused on the two 
sector case for simplicity, it would be interesting to complete a more 
thorough analysis of the case of multiple goldstini.  Finally, we hope 
to explore more fully the cosmological implications of this large class 
of theories.

\begin{acknowledgments}
We thank N.~Arkani-Hamed, A.~Arvanitaki, N.~Craig, S.~Dimopoulos, 
D.~Freedman, M.~Schmaltz, and D.~Shih for interesting discussions. 
The work of C.C. and Y.N. was supported in part by the Director, Office 
of Science, Office of High Energy and Nuclear Physics, of the US Department 
of Energy under Contract DE-AC02-05CH11231, and in part by the National 
Science Foundation under grants PHY-0555661 and PHY-0855653.  J.T. is 
supported by the U.S. Department of Energy under cooperative research 
agreement DE-FG0205ER41360.
\end{acknowledgments}

\appendix

\section{Explicit SUGRA Calculation}
\label{app:explicitcalc}

The goldstini mass spectrum derived in \Sec{sec:masses} can also be derived 
by explicit computation, using the SUGRA formalism of Ref.~\cite{Wess:1992cp}. 
The simplest case to consider is $N$ sequestered sectors labeled by $i$ 
that each contain only a single light chiral multiplet $X_i$.  That is, 
we assume that any other multiplets in sector $i$ have a supersymmetric 
mass term and can be integrated out of the effective SUGRA Lagrangian. 
In particular, this means that all moduli must be stabilized in the 
supersymmetric limit.

We start from a K\"{a}hler potential and superpotential of the sequestered 
form~\cite{Randall:1998uk}
\be
  K =  - 3 \MPl^2 \ln\left( \frac{-1}{3\MPl^2} 
    \sum_i \Omega^{(i)}(X_i, X_i^\dagger) \right),
\label{eq:Ksequester}
\ee
\be
  W = W_0 + \sum_i W^{(i)}(X_i),
\label{eq:Wsequester}
\ee
where each $\Omega^{(i)}$ and $W^{(i)}$ is only function of a single $X_i$. 
Here, $W_0$ is a constant that must be tuned to make the cosmological 
constant zero, and we can take $W_0$ to be real without loss of generality. 
It is convenient to define the modified K\"{a}hler potential
\be
  G = \frac{K}{\MPl^2} + \ln \frac{W}{\MPl^3} + \ln \frac{W^*}{\MPl^3},
\ee
and its derivatives $G_i = \partial_i G$, $G_{j^*} = \partial_{j^*} G$, 
$g_{ij^*} = \partial_i \partial_{j^*} G$, where
\be
  \partial_i \equiv \MPl \frac{\partial}{\partial X_i},
\qquad
  \partial_{j^*} \equiv \MPl \frac{\partial}{\partial X^\dagger_j}.
\ee
The K\"{a}hler metric $g_{ij^*}$ and its inverse $g^{ij^*} = (g^{-1})_{ji}$ 
can be used to raise and lower indices, such that $G^i = g^{ij^*} G_{j^*}$. 
With this notation, the scalar potential is
\be
  V = \MPl^4 \, e^G (G_i G^i - 3).
\ee
The condition for vanishing cosmological constant (and hence flat space) is
\be
G_i G^i = 3,
\label{eq:zerocosmo}
\ee
and the minimum of the potential satisfies
\be
  \partial_i V = 0,
\quad
  \partial_{j^*} V = 0.
\label{eq:minpotential}
\ee

After SUSY is broken, one linear combination of the fermionic components 
of $X_i$ is the true goldstino and is eaten to form the longitudinal 
component of the gravitino
\be
  \eta_{\rm long} = \frac{1}{\sqrt{3}} G_i \psi^i.
\ee
The gravitino mass is
\be
  m_{3/2} = \MPl \, e^{G/2}.
\ee
In unitary gauge, the remaining fermions have a quadratic Lagrangian of 
the form
\be
  -i \tilde{g}_{ij^*} \bar{\psi}^{j} \bar{\sigma}^\mu \partial_\mu \psi^i 
  - \frac{1}{2} m_{ij} \psi^i \psi^j 
  - \frac{1}{2} m^*_{i^*j^*} \bar{\psi}^i \bar{\psi}^j,
\ee
where $\tilde{g}$ is the K\"{a}hler metric with the true goldstino 
direction removed.  The mass matrix is
\be
  m_{ij} = m_{3/2} \left( \nabla_i G_j + \frac{1}{3} G_i G_j \right),
\ee
where $\nabla_i G_j = \partial_i G_j - \Gamma^k_{ij} G_j$ depends on 
the Christoffel symbol $\Gamma^k_{ij}$ derived from the K\"{a}hler 
metric.  Note that the direction corresponding to the eaten goldstino 
has a zero mass eigenvalue (assuming vanishing cosmological constant). 
The remaining $N-1$ uneaten goldstini masses can be determined by the 
physical mass-squared matrix
\be
  M^2 = A A^*,
\qquad
  {A_i}^{j^*} = m_{ik} g^{k j^*},
\ee
where $A^*$ is the complex conjugate of the matrix (not the Hermitian 
conjugate).  In $A$, it is possible to use $g$ instead of $\tilde{g}$ 
since the true goldstino direction is zeroed out by $m$.  Note that 
for the mass-squared matrix $M^2$ (unlike for $m$), we need not assume 
the $X_i$ have canonically normalized kinetic terms.

The key assumption of this paper is that SUSY is broken in the global 
limit $\MPl \rightarrow \infty$.  Moreover, we assume that any mixing 
between the chiral multiplets $X_i$ and the gravity multiplet is 
a subdominant effect, meaning that at the minimum of the potential
\be
  \epsilon_i \equiv \sqrt{ \frac{1}{3 \MPl^2} \frac{\partial_i \Omega\, 
    \partial_{i^*} \Omega}{\partial_i \partial_{i^*} \Omega}} \ll 1.
\ee
This corresponds to the assumption that there are no large linear terms 
in the K\"{a}hler potential, and in particular implies that Polonyi-like 
fields must have vevs $\langle X_i \rangle \ll \MPl$.

It is now a straightforward exercise to calculate the eigenvalues 
of $M^2$ as a series expansion in $\epsilon_i$.  Using 
\Eqs{eq:Ksequester}{eq:Wsequester}, one finds
\begin{align}
  {A_i}^{j^*} =&~~ {\delta_i}^{j^*} \left( 2m_{3/2} 
    + \frac{\partial_i V}{\partial_i W} \right) e^{2 i \theta_i}
\nonumber\\
  &~~ - \frac{2}{3} m_{3/2} G_i G^{j^*} + O(\epsilon_i),
\label{eq:finalmassmatrix}
\end{align}
where
\be
  \theta_i = \arg \left( \partial_i W \right).
\ee
By the condition in \Eq{eq:minpotential}, the $\partial_i V$ term in 
\Eq{eq:finalmassmatrix} vanishes, and because the uneaten goldstini are 
all orthogonal to $\eta_{\rm long}$, the $G_i G^{j^*}$ term is irrelevant. 
The $\theta_i$ phases in $A$ are also irrelevant, since $M^2 = A A^*$. 
So as advertised, one finds that the $N-1$ uneaten goldstini all have 
masses of $2m_{3/2}$ with corrections of order $\epsilon_i$.

One can also use the mass-squared matrix $M^2$ to calculate the eigenvalues 
for more general scenarios where $\epsilon_i$ is not small.  One amusing 
example is to consider $N-1$ sectors with $\epsilon_i \ll 1$, and an 
additional ``almost no-scale'' sector with arbitrary $\epsilon_{N}$ 
but $W^{(N)} = 0$.  In that case, one can show that of the $N-1$ goldstini, 
one is massless to all orders in $\epsilon_i$ (it only gets a mass 
proportional to $\partial_N W$).  The other $N-2$ goldstini get a mass
\be
  2 m_{3/2} \left( \frac{1}{1 + \epsilon_N^2} \right) + O(\epsilon_i).
\label{eq:massconformal}
\ee
Note that when $\epsilon_N = 0$, this reduces to the previous result, 
since in that limit $X_N$ is simply an extra massless mode that does 
not contribute to SUSY breaking.  We will explore these and other cases 
in future work.  As a preview, the result in \Eq{eq:massconformal} is 
equal to $2 F_C + O(\epsilon_i)$, where $F_C$ is the highest component 
of the conformal compensator.

\end{document}